\documentclass[twocolumn,showpacs,aps,prc,superscriptaddress,floatfix]{revtex4}

\usepackage[dvips]{graphicx}

\begin{document}

\title{
Measurement of the Generalized Polarizabilities of
the Proton in Virtual Compton Scattering at 
$Q^2$=0.92 and 1.76 GeV$^2$ }

\author{G.~Laveissi\`{e}re}
\affiliation{Universit\'{e} Blaise Pascal/IN2P3, F-63177 Aubi\`{e}re, France}
\author{L.~Todor}
\affiliation{Old Dominion University, Norfolk, VA 23529}
\author{N.~Degrande}
\affiliation{University of Gent, B-9000 Gent, Belgium}
\author{S.~Jaminion}
\affiliation{Universit\'{e} Blaise Pascal/IN2P3, F-63177 Aubi\`{e}re, France}
\author{C.~Jutier}
\affiliation{Universit\'{e} Blaise Pascal/IN2P3, F-63177 Aubi\`{e}re, France}
\affiliation{Old Dominion University, Norfolk, VA 23529}
\author{R.~Di Salvo}
\affiliation{Universit\'{e} Blaise Pascal/IN2P3, F-63177 Aubi\`{e}re, France}
\author{L.~Van Hoorebeke}
\affiliation{University of Gent, B-9000 Gent, Belgium}
\author{L.C.~Alexa}
\affiliation{University of Regina, Regina, SK S4S OA2, Canada}
\author{B.D.~Anderson}
\affiliation{Kent State University, Kent OH 44242}
\author{K.A.~Aniol}
\affiliation{California State University, Los Angeles, CA 90032}
% ...
\collaboration{The Jefferson Lab Hall A Collaboration}
\noaffiliation

%
%
%
%
%
%
\begin{abstract}
We report a Virtual Compton Scattering  study of the proton 
at low CM energies. We have determined the structure functions 
$P_{LL}-P_{TT}/\epsilon$ and $P_{LT}$, and the 
electric and magnetic Generalized Polarizabilities (GPs) 
$\alpha_E(Q^2)$ and $\beta_M(Q^2)$  
at momentum transfer Q$^2$= 0.92 and 1.76 GeV$^2$.
The electric GP shows a strong fall-off with $Q^2$ and its
global behavior does not follow a simple dipole form.
The magnetic GP shows a rise and then a fall-off;
this can be interpreted as the dominance of a long-distance
diamagnetic pion cloud at low $Q^2$, compensated at higher 
$Q^2$ by a paramagnetic contribution from $\pi N$ intermediate 
states.
\end{abstract}



\maketitle

%
%
%
%
%
%

%
%
%
%
%
%

The electric and magnetic polarizabilities of the nucleon
describe its response to a static electromagnetic field.  
Contrary to atomic polarizabilities, which are of the size
of the atomic volume~\cite{Dzuba:1997df},  
the proton electric polarizability $\alpha_E$~\cite{OlmosdeLeon:2001zn} 
is much smaller than one cubic fm, the volume scale of a nucleon.
Such a small polarizability is a natural indication of the
intrinsic relativistic character of the nucleon,
as illustrated in a harmonic oscillator model~\cite{Holstein:1999uu}. 
The smallness of the proton magnetic polarizability $\beta_M$
relative to $\alpha_E$ reflects a strong cancellation of 
para- and dia-magnetism in the proton.      

In Virtual Compton Scattering (VCS) \ $\gamma^* p \rightarrow \gamma p$ \
the polarizabilities become dependent on the momentum,
or the four-momentum transfer $Q^2$ of the virtual photon,
as first introduced by Guichon {\it et al.\/}~\cite{Guichon:1995pu}.
These Generalized Polarizabilities (GPs) can be seen
as Fourier transforms of local polarization densities
(electric, magnetic, and spin)~\cite{L'vov:2001fz}. 
Therefore they are a new  probe of the nucleon dynamics,
allowing e.g. to study the role of 
the pion cloud and quark core contributions to the nucleon GPs
at various length scales.
VCS can be accessed experimentally via exclusive
photon electroproduction  $ep \to ep \gamma$.
After the NE-18 experiment~\cite{vandenBrand:1995wj} and the
pioneering VCS experiment at MAMI~\cite{Roche:2000ng}, we performed
the E93-050 H$(e,e'p)\gamma$
experiment~\cite{Bertin:1993} at the Thomas Jefferson National Accelerator 
Facility (JLab).
We report low-energy expansion (LEX) analyses of our data 
up to pion threshold, and Dispersion Relation (DR) analyses of
our data extending into the $\Delta$-resonance region.


To lowest order in the fine structure constant $\alpha_{em}$,
the unpolarized 
$e p \rightarrow e p \gamma$ cross section at small $q'$ is:
\begin{eqnarray}
 d^5 \sigma ^{EXP} &=&   
 d^5 \sigma ^{BH+Born}  \ + \ q' \phi \Psi_0 \ + \ {\cal O } (q'^2) \ ,
 \nonumber \\ 
\Psi_0 &=& v_1 \cdot 
(P_{LL} - {\displaystyle 1 \over \displaystyle \epsilon} P_{TT}) 
\ + \ v_2 \cdot  P_{LT}  
\label{eq01} 
\end{eqnarray} 
where $\phi, v_1, v_2$ are kinematical coefficients
defined in~\cite{Guichon:1998xv}, $q'$ is the final photon energy in 
the $\gamma p$ CM frame, 
and $\epsilon$ is the virtual photon polarization.
$d^5 \sigma ^{BH+Born}$
corresponds to the coherent sum of the Bethe-Heitler (BH)
and the VCS Born amplitudes, and
depends only on the
elastic form factors $G_{E}^p, G_{M}^p$ of the proton.  
This is a particular case
of Low's low-energy theorem~\cite{Low:1958sn} 
for threshold photon production. The structure functions:
\begin{eqnarray}
P_{LL}-{1 \over \epsilon} P_{TT}  =  
{ 4 M_p \over \alpha_{em} } 
G_E^p(Q^2) \alpha_E (Q^2) + \mbox{[spin-flip GPs]}  \nonumber  \\
P_{LT} =  - { 2 M_p \over \alpha_{em} } 
 \sqrt{ {q^2 \over Q^2} } 
G_E^p(Q^2) \beta_M (Q^2) + \mbox{[spin-flip GPs]}  \label{eq02} 
\end{eqnarray} 
contain five of the six independent 
GPs~\cite{Drechsel:1997ag,Drechsel:1998xv}. 
These structure functions are defined at fixed $q$, 
the CM three-momentum of the VCS virtual photon.
Equivalently, $Q^2$ in Eqs.(\ref{eq02}) 
is defined in the $q'\rightarrow 0$ limit:
$Q^2 = 2 M_p \cdot ( \sqrt{M_p^2 + { q}^{ 2}} - M_p)$.

%
%
%
%
%
%
The apparatus, running conditions and analyses of the 
JLab experiment are detailed 
elsewhere~\cite{Laveissiere:2003jf,Degrande:2001th,
Jaminion:2001th,Jutier:2001th,Laveissiere:2001th,Todor:2000th}. 
An electron beam of 4.030 GeV energy 
was directed onto a 15 cm liquid hydrogen target.
The two Hall~A Spectrometers were used to 
detect the scattered electron and the outgoing proton in coincidence,
allowing the identification of the exclusive 
reaction $ ep \to ep \gamma$ \ by  the 
missing-mass technique.
This experiment makes use of the full capabilities of the 
accelerator and the Hall~A instrumentation~\cite{Alcorn:2003}:
100\% duty cycle, high resolution spectrometers, 
high luminosities.
We summarize our kinematics in Table~\ref{datasets}.
Variables such as 
$q'$, or the CM polar and azimuthal angles $ \theta$ and
$\varphi$ of the outgoing photon w.r.t. $\vec q$, 
are obtained by reconstructing the missing particle. 
The acceptance calculation is provided by 
a dedicated Monte-Carlo simulation~\cite{VanHoorebeke} 
including a model cross section, resolution effects 
and radiative corrections~\cite{Vanderhaeghen:2000ws}.
A number of cuts are applied in event analysis, especially to
obtain a well-defined acceptance and to
eliminate protons punching through the spectrometer
entrance collimator.

\begin{table}[hb]
\caption{\label{datasets} Kinematics of $ep\rightarrow e p \gamma$.
We used data sets I-a and II for the LEX analyses and all data sets
 for the DR analyses.
 }
\begin{ruledtabular}
\begin{tabular}{ccc} 
data set & $Q^2$-range (GeV$^2$) & $W$-range   \\
\hline
I-a & [0.85, 1.15] & mostly $< \pi N$ threshold \\
I-b & [0.85, 1.15] & mostly $\Delta(1232)$ resonance \\
II  & [1.60, 2.10] & mostly $< \pi N$ threshold \\
\end{tabular}
\end{ruledtabular}
\end{table}

\begin{figure}[hb]
\includegraphics[width=8.6cm,height=5.0cm]{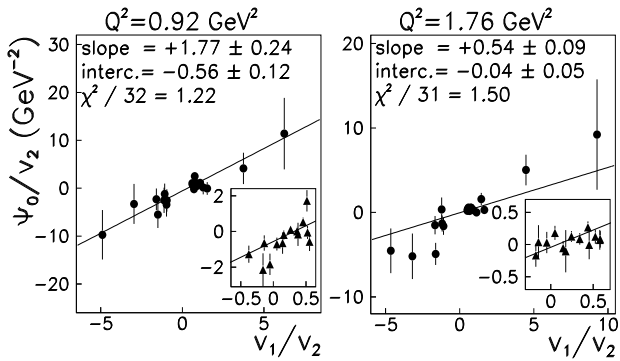}
\caption{\label{fitgps}
A graphical representation of the LEX fit (straight line) 
for data sets I-a and II. 
Circles correspond to out-of-plane data,
and the inner plot is a zoom on the lepton plane
data (triangles). $\Psi_0, v_1$ and $v_2$ are defined in the text. }
\end{figure}

%
%
%
%
%
%
We performed LEX analyses of the data sets I-a and II.
The photon electroproduction cross section 
is determined 
as a function of $q', \theta$ and $\varphi$ 
at a fixed value of $q$ (1.080 and 1.625 GeV/c)
and $\epsilon$ (0.95 and 0.88, respectively).
The effect of the GPs on the cross section is small, 
reaching at maximum 10-15\% below pion threshold. 
The method to extract the structure functions 
is deduced from Eq.(\ref{eq01}), in which 
the (BH+Born) cross section is calculated using
a recent parametrization of the 
proton form factors~\cite{Brash:2001qq}.
For each bin in $(\theta, \varphi)$, we measure 
$d^5 \sigma ^{EXP}$ in several bins in $q'$,
and extrapolate the quantity 
$\Delta {\cal M} = 
 (d^5 \sigma ^{EXP} -  d^5 \sigma ^{BH+Born})/( \phi q' )$
to $q'=0$, yielding the value of $\Psi_0$. 
In our data, $\Delta {\cal M}$ does not exhibit 
any significant $q'$-dependence, so the extrapolation to $q'=0$
is done in each bin in $(\theta, \varphi)$ 
by averaging $\Delta {\cal M}$ over $q'$.
The resulting $\Psi_0$ term is then fitted as a linear 
combination of two free parameters, which are 
the structure functions $P_{LL}-P_{TT}/\epsilon$ and 
$P_{LT}$ (Fig.~\ref{fitgps}).

The systematic errors 
are calculated from four sources added quadratically:
1) $\pm$ 2 MeV uncertainty in beam energy, 
2) $\pm$ 0.5 mrad uncertainty in horizontal angles, 
3) $\pm$ 2.3\% uncertainty in overall absolute cross section normalization, 
and 4) $\pm$ 2\% uncertainty due to possible cross section shape distortions.
The value of the reduced $\chi^2$ of the fit
(Fig.~\ref{fitgps}) is one measure of the validity
of the LEX in our kinematics.
The LEX results for the structure functions are summarized in 
Table~\ref{Results}.

%
%
%
%
%
%
We performed DR analyses of the data sets I-a and II, and also I-b
including data from
$\pi N$ threshold through the $\Delta$ resonance.
In the DR formalism of Pasquini {\it et al.}~\cite{Pasquini:2001yy},
the VCS amplitude is determined by unitarity
from the MAID $\gamma^{(\ast)}p\rightarrow N\pi$ 
multipoles~\cite{Drechsel:1998hk},  plus asymptotic terms 
$\Delta\alpha$, $\Delta\beta$
which
are unconstrained phenomenological contributions to the GPs
$\alpha_E(Q^2)$ and $\beta_M(Q^2)$.
$\Delta\alpha$, $\Delta\beta$ are parametrized as follows:
\begin{eqnarray}
\Delta\alpha(Q^2)=
\alpha_E(Q^2) - \alpha_E^{\pi N}(Q^2)= 
{ \displaystyle [ \alpha_{E}^{exp}   -  
\alpha_{E}^{\pi N} ]_{Q^2=0} 
\over
\displaystyle ( 1 + Q^2/ \Lambda_{\alpha}^2 )^2 }
\label{eq03}
\end{eqnarray}
(same relation for $\Delta\beta$ with parameter $\Lambda _{\beta}$) 
where $\alpha_E^{\pi N}$ ($\beta_M^{\pi N}$)
is the $\pi N$ dispersive contribution evaluated from MAID, 
 $\alpha_E^{exp}$ ($\beta_M^{exp}$) is 
the experimental value at $Q^2=0$~\cite{OlmosdeLeon:2001zn},
and the mass coefficients 
$\Lambda _{\alpha}$ and $\Lambda _{\beta}$ 
are free parameters. Theoretically, the choice of the 
dipole form in Eq.(\ref{eq03}) is not compulsory.
More fundamentally, the DR model
provides a rigorous treatment of the higher order terms
in the VCS amplitude up to the $N \pi \pi$ threshold, by
including resonances in the $\pi N$ channel.
In the region of the $\Delta(1232)$ resonance, 
these higher order terms become dominant over
the lowest order GPs given by the LEX.

\begin{table}[ht]
\caption{\label{Results} Compilation of 
the VCS structure functions. 
In all cases the first error is statistical, 
and the second one is the total systematic error.
 }
\begin{ruledtabular}
\begin{tabular}{ccccc}
\ & $Q^2$  & $\epsilon$  &
$P_{LL}-P_{TT}/\epsilon$ & $P_{LT}$ \\
\multicolumn{2}{r}{(GeV$^2$)}  & \ & (GeV$^{-2}$) & (GeV$^{-2}$) \\
 \hline
Ref. & \multicolumn{4}{c}{Previous experiments} \\
\hline
\cite{OlmosdeLeon:2001zn} & 0 &  \ & 81.3 $\pm \, 2.0 \pm 3.4$ & -5.4 $\pm \, 1.3 \pm 1.9$ \\
\cite{Roche:2000ng} & 0.33 & 0.62  & 23.7 $\pm \, 2.2 \pm 4.3$ & -5.0 $\pm \, 0.8 \pm 1.8$ \\
\hline
Set & \multicolumn{4}{c}{ This experiment, LEX Analyses}  \\ 
\hline
I-a  & 0.92 & 0.95 & \textbf{1.77} $\pm \, 0.24 \pm 0.70$ & \textbf{-0.56} $\pm \, 0.12 \pm 0.17$ \\
II   & 1.76 & 0.88 & \textbf{0.54} $\pm \, 0.09 \pm 0.20$ & \textbf{-0.04} $\pm \, 0.05 \pm 0.06$ \\
 \hline
Set & \multicolumn{4}{c}{ This experiment, DR Analyses}  \\ 
 \hline
I-a  & 0.92 & 0.95 & \textbf{1.70} $\pm \, 0.21 \pm 0.89$ & \textbf{-0.36} $\pm \, 0.10 \pm 0.27$ \\
I-b  & 0.92 & 0.95 & \textbf{1.50} $\pm \, 0.18 \pm 0.19$ & \textbf{-0.71} $\pm \, 0.07 \pm 0.05$ \\
II   & 1.76 & 0.88 & \textbf{0.40} $\pm \, 0.05 \pm 0.16$ & \textbf{-0.09} $\pm \, 0.02 \pm 0.03$ \\
\end{tabular}
\end{ruledtabular}
\end{table}

The DR analysis consists in fitting the free parameters 
$\Lambda_{\alpha}$ and $\Lambda_{\beta}$ to our cross-section 
data. This yields the value of the GPs  
$\alpha_E(Q^2)$ and $\beta_M(Q^2)$ using Eq.(\ref{eq03}).
This also yields the value of the structure functions
of Eqs.(\ref{eq02}) since the DR model predicts all the
spin-flip GPs~\cite{Pasquini:2001yy}. Our DR results are 
presented in Tables~\ref{Results} and \ref{resultslalb}.
The systematic uncertainties are calculated
from the same sources as in the LEX analyses.
The error bars differ from one data set to another, 
due to differences in
phase space coverage and in sensitivity to both the
physics and the sources of systematic errors.
The reasonably good $\chi^2$ of the DR fits (1.3 to 1.5) 
indicates that the DR model allows 
a reliable extraction of GPs in our kinematics, 
both below and above pion threshold.

%
%
%
%
%
%

\begin{table}[b]
\caption{\label{resultslalb} 
The dipole mass parameters $\Lambda_{\alpha}$ and $\Lambda_{\beta}$
obtained by fitting the three data sets
independently, and the electric and magnetic GPs evaluated
at $Q^2$= 0.92 GeV$^2$ (data sets I-a, I-b) and 1.76 GeV$^2$
(data set II).
The first and second errors are 
statistical and total systematic errors, respectively.
}
\begin{ruledtabular}
\begin{tabular}{ccc}
 data set
& $\Lambda_{\alpha}$ (GeV) & $\Lambda_{\beta}$  (GeV) \\
\hline
I-a & 0.741 $\pm \, 0.040 \pm 0.175$ & 0.788  $\pm \, 0.041 \pm 0.114$  \\ 
I-b & 0.702 $\pm \, 0.035 \pm 0.037$ & 0.632  $\pm \, 0.036 \pm 0.023$  \\
II &  0.774 $\pm \, 0.050 \pm 0.149$ & 0.698  $\pm \, 0.042 \pm 0.077$  \\
\hline
data set   & $\alpha_E(Q^2)$  ($10^{-4}$ fm$^3$) 
& $\beta_M(Q^2)$  ($10^{-4}$ fm$^3$) \\
 \hline
I-a         & 1.02  $\pm \, 0.18 \pm 0.77$    
            & 0.13  $\pm \, 0.15 \pm 0.42$  \\ 
I-b         & 0.85  $\pm \, 0.15 \pm 0.16$ 
            & 0.66  $\pm \, 0.11 \pm 0.07$  \\
II          & 0.52  $\pm \, 0.12 \pm 0.35$
            & 0.10  $\pm \, 0.07 \pm 0.12$   \\
\end{tabular}
\end{ruledtabular}
\end{table}

\begin{figure}[ht]
\includegraphics[width=8.6cm]{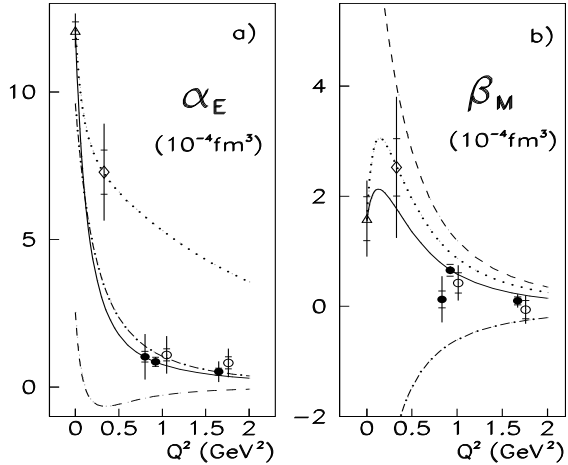}
\caption{\label{plotsf}
Compilation of the data on electric (a) and
magnetic (b) GPs. 
Data points are from Ref.~\cite{OlmosdeLeon:2001zn}
({\tiny $\bigtriangleup$}), the LEX analysis of 
MAMI~\cite{Roche:2000ng} ($\diamond$),  and the present
LEX ($\circ$) and DR ($\bullet$) analyses of JLab.
Some JLab points are shifted in abscissa for better visibility.
The inner error bar is statistical; the outer one is the total
error (statistical plus systematic). 
The curves show calculations in the DR model (see text).
}
\end{figure}

Figure ~\ref{plotsf} shows our DR extraction of the 
GPs $\alpha_E(Q^2)$ and $\beta_M(Q^2)$,
together with the point at $Q^2$= 0~\cite{OlmosdeLeon:2001zn}
and the points derived from LEX analyses.
The latter are obtained by subtracting the spin-flip 
polarizability predictions~\cite{Pasquini:2001yy} to 
the structure functions of Eqs.(\ref{eq02}).
This involves some model dependence, which is not presently
taken into account in the error bars.

The solid curves in Fig.~\ref{plotsf} 
are the full DR calculations,
split into their dispersive $\pi N$ contributions (dashed)
and the remaining asymptotic contributions of Eq.(\ref{eq03})
(dash-dotted) for
$\Lambda_{\alpha}$=0.70 GeV and $\Lambda_{\beta}$=0.63 GeV,
as fitted to the JLab data set I-b.
The $\pi N$ contribution to the magnetic polarizability in 
Fig.~\ref{plotsf}-b is strongly paramagnetic,
predominantly arising from the $\Delta(1232)$ resonance. 
In the DR formalism, 
this is cancelled by a strong diamagnetic term $\Delta \beta$
originating from $\sigma$-meson $t$-channel exchange.
The interpretation of $\Delta \beta$ as the contribution of a 
long-distance pion cloud is further supported by the fact 
that the fitted scale parameter $\Lambda_{\beta}$=0.63 GeV is
smaller than the elastic form factor dipole parameter 
$\Lambda$=0.84 GeV.
The dotted curves in Fig.~\ref{plotsf} result from the
full DR calculation, evaluated with
$\Lambda_{\alpha}$=1.79 GeV and $\Lambda_{\beta}$=0.51 GeV,
which reproduces the MAMI LEX data.
The data for $\alpha_E(Q^2)$ disagree strongly with the simple dipole
ansatz for the contribution $\Delta \alpha$.
It should be noted that our DR analysis is basically insensitive to
the particular choice of form of $\Delta \alpha$ and $\Delta \beta$,
since our fits are performed independently in two small ranges of $Q^2$.
Finally we point out that
the $\eta N$ and $\pi\pi N$ channels, which must contribute to
$\Delta \alpha$, have
resonances ($S_{11}(1535)$ and $D_{13}(1520)$, respectively) with
transition form factors that do not follow
a simple dipole $Q^2$ dependence~\cite{Thompson:2000by,Tiator:2003uu}.

%
%
%
%
%
%
%
%
In summary, we studied the process $e p \to e p \gamma$ at JLab. 
With data below pion threshold we applied the LEX, and 
for data extending through the $\Delta$ resonance we applied the
DR formalism to extract the Generalized Polarizabilities.
The different analyses are consistent, and the results give new insight
into the correlations between spatial and dynamical variables
in the proton. Other experiments at low energy will measure 
the VCS structure functions 
at low $Q^2$~\cite{Shaw:1997,Merkel:2000} and separate the
six GPs via double polarization 
measurements~\cite{Merkel:2000,Vanderhaeghen:1997bx}.


We thank the JLab accelerator staff and the Hall~A technical staff
for their dedication.
This work was supported by DOE contract DE-AC05-84ER40150 under
which the Southeastern Universities Research Association (SURA)
operates the Thomas Jefferson National Accelerator Facility. We
acknowledge additional grants from the US DOE and NSF, the French
CNRS and CEA,
 the Conseil R\'egional d'Auvergne, the
FWO-Flanders (Belgium) and the BOF-Gent University. 
We thank the INT (Seattle) and ECT* (Trento) for 
the organization of VCS workshops.

\bibliography{common}

\end{document}